# Phase Diagram for the *O(n)* Model with Defects of "Random Local Field" Type and Verity of the Imry-Ma Theorem

*A.A. Berzin¹, A.I. Morosov²\*, and A.S. Sigov¹*

¹ Moscow Technological University (MIREA), 78 Vernadskiy Ave., 119454 Moscow, Russian Federation

² Moscow Institute of Physics and Technology (State University), 9 Institutskiy per., 141700 Dolgoprudny, Moscow Region, Russian Federation

## Abstract

It is shown that the Imry-Ma theorem stating that in space dimensions $d<4$ the introduction of an arbitrarily small concentration of defects of the "random local field" type in a system with continuous symmetry of the *n*-component vector order parameter (*O(n)* model) leads to the long-range order collapse and to the occurrence of a disordered state, is not true if the anisotropic distribution of the defect-induced random local field directions in the *n*-dimensional space of the order parameter leads to the defect-induced effective anisotropy of the "easy axis" type. For a weakly anisotropic field distribution, in space dimensions $2 \leq d < 4$ there exists some critical defect concentration, above which the inhomogeneous Imry-Ma state can exist as an equilibrium one. At lower defect concentration the long-range order takes place in the system. For a strongly anisotropic field distribution, the Imry-Ma state is suppressed completely and the long-range order state takes place at any defect concentration.

\*E-mail: mor-alexandr@yandex.ru

## I. Introduction

After the publication in 1975 the classical paper by Imry and Ma [1], the viewpoint was firmly established in the literature that at space dimensions $d < 4$ the introduction of an arbitrarily small concentration of defects of the "random local field" type in a system with continuous symmetry of the $n$-component vector order parameter ($O(n)$ model) leads to the long-range order collapse and to the occurrence of a disordered state, which in what follows will be designated as the Imry-Ma state and the statement given above will be named the Imry-Ma theorem.

The proof of this theorem is based on simple energy considerations. Let in each area of the investigated system with a linear size $L$ the order parameter line up along the total field created by defects in this area due to predominance of one direction of the field because of concentration fluctuations. Thus the gain in the volume-energy density comprises the value of $w_{def} \propto L^{-d/2}$.

In different areas of the system the total field directions differ. It causes the order parameter inhomogeneity on the spatial scale of $L$. A loss in the exchange energy volume density due to inhomogeneity of the order parameter constitutes the value of $w_{ex} \propto L^{-2}$.

It can be easily seen that at $d<4$ and for big $L$ values the energy gain exceeds the energy loss. Hence the inhomogeneous state inevitably occurs in space dimensions $d<4$.

It should be mentioned that the above consideration holds only for isotropic systems. The presence of weak anisotropy of the "easy axis" type leads to the fact that the order parameter degradation and the occurrence of the Imry-Ma inhomogeneous state take place if the defect concentration exceeds some critical value [2].

This notice has seemingly nothing to do with initially isotropic $O(n)$-model. However it was shown in our recent paper [3] that an anisotropic distribution of the directions of defect-induced random local fields in the order parameter *n*-dimensional space gives rise to the effective anisotropy in the system. For a qualitative explanation of the reasons for the occurrence of the effective anisotropy, let us consider the influence of local field induced by *l*-th defect $\mathbf{h}_l$ upon uniform distribution of the order parameter. For simplicity, we neglect the longitudinal susceptibility of the system at low temperatures, much smaller than the temperature of magnetic ordering.

The random field component $\mathbf{h}_l^\perp$ perpendicular to the order parameter direction $\mathbf{s}_0$ in the system free of defects leads to a local deviation of the order parameter and to the appearance of the component $\mathbf{s}^\perp(\mathbf{r})$ orthogonal to $\mathbf{s}_0$. The result is a negative additive to the energy of the ground state proportional to $(\mathbf{h}_l^\perp)^2$. It is maximum in modulus when the vector $\mathbf{s}_0$ direction is perpendicular to the defect-induced local field. Consequently the system benefits if the order parameter is oriented perpendicularly to the predominant direction of random fields.

Thus the Imry-Ma theorem is obviously valid only for ideally isotropic distribution of random local field directions, while the case of anisotropic field distribution needs a detailed analysis to be considered in the present paper.

## II. Energy of a system of classical spins

The exchange-interaction energy of *n*-component localized unit spins $\mathbf{s}_i$ (a vector length can be included in corresponding interaction or field constants) comprising the simple cubic *d*-dimensional lattice, within the nearest-neighbors approximation has the form

$$W_{ex} = -\frac{1}{2} J \sum_{i,\delta} \mathbf{s}_i \mathbf{s}_{i+\delta}, \tag{1}$$

where *J* is the exchange interaction constant, the summation in *i* is performed over the whole spin lattice, and the summation in $\delta$ is performed over the nearest neighbors.

The energy of interaction between the spins and defect-induced random local fields is

$$W_{def} = -\sum_l \mathbf{s}_l \mathbf{h}_l, \tag{2}$$

where the summation is performed over defects randomly located in the lattice sites, and the density of random local fields $\mathbf{h}$ distribution in the spin space (order parameter space) possesses the property $\rho(\mathbf{h}) = \rho(-\mathbf{h})$, which provides the lack of mean field in an infinite system.

Let the net energy $W = W_{ex} + W_{def}$ be expanded as a power series in $\mathbf{s}^\perp(\mathbf{r})$. The order parameter component $s_i^\parallel$ parallel to the direction of its mean quantity is

$$s_i^\parallel = \sqrt{1 - (\mathbf{s}_i^\perp)^2}. \tag{3}$$

Then one has

$$\mathbf{s}_i \mathbf{s}_{i+\delta} = \sqrt{(1 - (\mathbf{s}_i^\perp)^2)\left(1 - \left(\mathbf{s}_{i+\delta}^\perp\right)^2\right)} + \mathbf{s}_i^\perp \mathbf{s}_{i+\delta}^\perp. \tag{4}$$

The quantity $\mathbf{s}_{i+\delta}^\perp$ can be expressed through the value $\mathbf{s}_i^\perp$, by expanding the function $\mathbf{s}^\perp(\mathbf{r})$ in the continuous representation in a Taylor series. Let us take up the question of neglecting the terms of this series that contain higher-order derivatives. Such a neglect is valid if the characteristic scale of the order parameter distortions far exceeds the interaction radius (i.e. interatomic distance in our case). As shown by a detailed examination, the distortions induced by random defect fields meet this condition for a two-dimensional ($d=2$) space and do not meet for $2<d<4$. In the latter case, higher derivatives renormalize only the value but not the angle and concentration dependences of the anisotropy energy. Therefore, to estimate the anisotropy energy by the order of magnitude we restrict ourselves to the terms containing the lowest degree of space derivatives.

The first two terms of the expansion of inhomogeneous exchange energy in powers of $\mathbf{s}^\perp(\mathbf{r})$ with regard to Eqs. (1) and (4) have the form

$$W_{ex} = \frac{D}{2}\int d^d\mathbf{r}\, \frac{\partial s^\perp}{\partial x_i}\frac{\partial s^\perp}{\partial x_i} + \frac{D}{8}\int d^d\mathbf{r}\,[\nabla(\mathbf{s}^\perp(\mathbf{r}))^2]^2, \tag{5}$$

where $b$ is the interstitial distance and $D = Jb^{2-d}$.

To the continuous approximation, the energy of random field $\mathbf{h}(\mathbf{r})$ interaction with the order parameter $\mathbf{s}(\mathbf{r})$ is

$$W_{def} = -\int d^d\mathbf{r}\, \mathbf{h}(\mathbf{r})\,\mathbf{s}(\mathbf{r}), \tag{6}$$

where

$$\mathbf{h}(\mathbf{r}) = \sum_l \mathbf{h}_l\, \delta(\mathbf{r} - \mathbf{r}_l). \tag{7}$$

The energy of random field $\mathbf{h}(\mathbf{r})$ interaction with the order parameter longitudinal component equals zero in view of the equality $\rho(\mathbf{h}) = \rho(-\mathbf{h})$. Therefore $W_{def}$ can be written as

$$W_{def} = -b^{-d}\int d^d\mathbf{r}\, \mathbf{h}^\perp(r)\,\mathbf{s}^\perp(r), \tag{8}$$

where

$$\mathbf{h}^\perp(\mathbf{r}) = b^d \sum_l [\mathbf{h}_l - \mathbf{s}_0(\mathbf{s}_0\mathbf{h}_l)]\,\delta(\mathbf{r} - \mathbf{r}_l). \tag{9}$$

### III. Effective anisotropy. Approximation quadratic in $\mathbf{h}^\perp(\mathbf{r})$

#### A. 2<d<4

In the given approximation, one can keep only the first term on the right side of Eq. (5) for inhomogeneous exchange energy. The Green function of such a problem is well known [1].

The Fourier component of the function $\mathbf{s}^\perp(\mathbf{k})$ is related to the Fourier component of the random field $\mathbf{h}^\perp(\mathbf{k})$

$$\mathbf{s}^\perp(\mathbf{k}) = \chi^\perp(\mathbf{k})\mathbf{h}^\perp(\mathbf{k}), \tag{10}$$

where the Fourier component of the corresponding susceptibility is

$$\chi^\perp(\mathbf{k}) = (Jb^2k^2)^{-1}, \tag{11}$$

and

$$\mathbf{h}^\perp(\mathbf{k}) = \frac{1}{V}\int d^d\mathbf{r}\, \mathbf{h}^\perp(\mathbf{r})\exp(-i\mathbf{k}\mathbf{r}) = \frac{1}{N}\sum_l[\mathbf{h}_l - \mathbf{s}_0(\mathbf{s}_0\mathbf{h}_l)]\exp(-i\mathbf{k}\mathbf{r}_l), \quad (12)$$

$V$ is the volume of the system, and $N$ is the number of elementary cells. Then

$$\mathbf{s}^\perp(\mathbf{r}) = \frac{1}{N}\sum_\mathbf{k}\chi^\perp(\mathbf{k})\sum_l[\mathbf{h}_l - \mathbf{s}_0(\mathbf{s}_0\mathbf{h}_l)]\exp[i\mathbf{k}(\mathbf{r} - \mathbf{r}_l)], \quad (13)$$

the summation is performed over all $\mathbf{k}$ from the Brillouin zone. Substitution of this expression into Eq. (8) gives

$$W_{def} = -\frac{1}{N}\sum_\mathbf{k}\chi^\perp(\mathbf{k})\sum_{l,m}[\mathbf{h}_l - \mathbf{s}_0(\mathbf{s}_0\mathbf{h}_l)][\mathbf{h}_m - \mathbf{s}_0(\mathbf{s}_0\mathbf{h}_m)]$$

$$\times \exp[i\mathbf{k}(\mathbf{r}_m - \mathbf{r}_l)]. \quad (14)$$

By virtue of a random distribution of defects in the coordinate space and a random choice of defect-induced local fields, a nonzero contribution to $W_{def}$ results from the summands with $l = m$. Therefore Eq. (14) yields

$$W_{def} = -\frac{1}{N}\sum_\mathbf{k}\chi^\perp(\mathbf{k})\sum_l[\mathbf{h}_l - \mathbf{s}_0(\mathbf{s}_0\mathbf{h}_l)]^2$$

$$= -x\sum_\mathbf{k}\chi^\perp(\mathbf{k})\langle[\mathbf{h}_l - \mathbf{s}_0(\mathbf{s}_0\mathbf{h}_l)]^2\rangle, \quad (15)$$

where $x$ is the dimensionless concentration of defects (the number of defects per a unit cell), and the brackets $\langle\ \rangle$ denote averaging over defect-induced local fields. Going from summation over $\mathbf{k}$ to integration over the Brillouin zone and introducing the notation

$$\tilde{\chi}^\perp = \int\frac{d^d\mathbf{k}}{(2\pi)^d}\chi^\perp(\mathbf{k}), \quad (16)$$

we get the volume density of energy of interaction between the order parameter and defect-induced random local fields

$$w_{def} = -x\tilde{\chi}^\perp[\langle\mathbf{h}_l^2\rangle - \langle(\mathbf{s}_0\mathbf{h}_l)^2\rangle]. \quad (17)$$

For space dimensions $2<d<4$, the quantity $\tilde{\chi}^\perp$ has no peculiarities at $\mathbf{k} = 0$.

In the case of anisotropic distribution of the directions of random fields, the second summand in square brackets in the right-hand side of Eq. (17) is "responsible" for anisotropy in the order parameter space. The inhomogeneous exchange energy contribution to the volume density of anisotropy energy can be found in a similar manner. It is of opposite sign and its magnitude is twice less than that of $w_{def}$ (17). Hence the resulting volume density of the anisotropy energy takes the form

$$w_{an} = \frac{x\tilde{\chi}^\perp}{2}[s_{01}^2\langle h_{l1}^2\rangle + s_{02}^2\langle h_{l2}^2\rangle + \ldots + s_{0n}^2\langle h_{ln}^2\rangle], \quad (18)$$

where $h_{lj}$ и $s_{0j}$ are $j$-th components of the field $\mathbf{h}_l$ and the order parameter $\mathbf{s}_0$ correspondingly.

The following value is taken as an effective anisotropy constant $K_{eff}$

$$K_{eff} = 2b^d(w_{an}^{max} - w_{an}^{min}), \quad (19)$$

where $w_{an}^{max}$ and $w_{an}^{min}$ are maximum and minimum values of $w_{an}$ as a function of vector $\mathbf{s}_0$ direction.

### B. $d=2$

A specific feature of two-dimensional models is the absence of the long-range order in a pure system at finite temperature and so one has to anticipate the existence of the long-range order induced by random fields and solve a self-consistent problem [4].

Since under the influence of random field the order parameter deviates from the easy direction to the hard one, the expression for $\chi^\perp(\mathbf{k})$ takes the form

$$\chi^\perp(\mathbf{k}) = (Jb^2k^2 + K_{eff})^{-1}. \quad (20)$$

It can be easily seen that $K_{eff}$ cuts the divergence of $\tilde{\chi}^\perp$ at small $\mathbf{k}$ values which bring the main contribution to $\tilde{\chi}^\perp$ for $d = 2$. As the result we obtain

$$\tilde{\chi}^\perp = \frac{1}{4\pi b^2 J}\ln\frac{4\pi J}{K_{eff}}. \quad (21)$$

The value of $K_{eff}$ can be found by solving the self-consistency equation (19) after substituting the value of $\tilde{\chi}^\perp$.

### C. Examples of distribution of random field directions

In a particular case of anisotropic distribution of random field directions when all $\mathbf{h}_l$ vectors are collinear, the order parameter is energetically favorable to be oriented perpendicularly to this direction. Thus in the case of the *X-Y* model (*n*=2), the global anisotropy of the "easy axis" type arises in the system, and in the case of the Heisenberg model (*n*=3), the global anisotropy of the "easy plane" type is induced by the defects. The volume density of the anisotropy energy takes the form

$$w_{an} = \frac{1}{2}x\tilde{\chi}^\perp \langle \mathbf{h}_l^2 \rangle \cos^2\varphi \equiv \frac{1}{2}K_{eff}b^{-d}\cos^2\varphi, \qquad (22)$$

where $\varphi$ is the angle between the order parameter vector and the axis of "hard magnetization" (defect-induced random fields are collinear to this axis).

In the case of coplanar and isotropic in the selected plane distribution of random fields in the Heisenberg model, the volume density of the anisotropy energy is

$$w_{an} = -\frac{1}{4}x\tilde{\chi}^\perp \langle \mathbf{h}_l^2 \rangle \cos^2\varphi \equiv -\frac{1}{2}K_{eff}b^{-d}\cos^2\varphi, \qquad (23)$$

$\varphi$ being the angle between the order parameter vector and the normal to the plane containing random field vectors, that is the "easy axis" type anisotropy takes place.

Evaluating the effective anisotropy constant $K_{eff}$ by the order of magnitude, one gets the quantity

$$K_{eff} \sim \tilde{\chi}^\perp b^d x \langle \mathbf{h}_l^2 \rangle. \qquad (24)$$

For $d = 3$ one has $\tilde{\chi}^\perp b^d J \sim 0.2$, and for $d = 2$ the solution of the self-consistency equation (19) for the collinear distribution of random fields, to a first approximation, yields

$$K_{eff} = \frac{x\langle \mathbf{h}_l^2 \rangle}{4\pi J} \ln \frac{16\pi^2 J^2}{x\langle \mathbf{h}_l^2 \rangle}. \qquad (25)$$

It should be mentioned that it is just the occurrence of the "easy type" effective anisotropy in the space with $d = 2$ which leads to the long-range ordering at finite temperature. Such a phenomenon was discovered under theoretical analysis of the two-dimensional *X-Y* model [5]. Since a long-range order is absent in a pure system at finite temperature, and the Berezinskii-Kosterlitz-Thouless phase takes place [6, 7], this phenomenon has been further named the "random fields induced order" (RFIO) [8]. In Ref. [8] this phenomenon was generalized to the Heisenberg model. As the reason for RFIO occurrence, the violation of the continuous symmetry of the system was indicated, while the microscopic mechanism of RFIO has been found in our paper [4].

Due to initiation of the weak anisotropy of the "easy axis" type the *X-Y* and Heisenberg models are transformed to the class of Ising models [9], that explains the appearance of the long-range order at finite temperature [5].

The defect-induced anisotropy of the "easy plane" type leads to transformation of the Heisenberg model to the class of *X-Y* models, and thus the Berezinskii-Kosterlitz-Thouless transition occurs in the system [9].

In the case of more general ellipsoidal distribution of random fields in the *X-Y* model: $|\mathbf{h}_l| = \text{const}$,

$$\rho(\mathbf{h}) = A[h_x^2 + (1+\varepsilon)h_y^2], \tag{26}$$

where constant $A$ can be found from the normalization condition and $\varepsilon > -1$, the quantity $K_{eff}$ given by Eq. (19) becomes

$$K_{eff} = \frac{|\varepsilon|\tilde{\chi}^\perp b^d x \langle \mathbf{h}_l^2 \rangle}{2(2+\varepsilon)}. \tag{27}$$

For both signs of $\varepsilon$, the defects induce anisotropy of the "easy axis" type, its direction being perpendicular to the predominant direction of random fields. For the Heisenberg model with the distribution: $|\mathbf{h}_l| = \text{const}$,

$$\rho(\mathbf{h}) = A[h_x^2 + h_y^2 + (1+\varepsilon)h_z^2], \tag{28}$$

the value of $K_{eff}$ is

$$K_{eff} = \frac{2|\varepsilon|\tilde{\chi}^{\perp} b^d x \langle \mathbf{h}_l^2 \rangle}{5(3+\varepsilon)}. \tag{29}$$

At $-1 < \varepsilon < 0$, the easy axis $z$ arises in the system, and at $\varepsilon > 0$, the easy plane $xy$ does exist.

### IV. Effective anisotropy of the fourth order in $\mathbf{h}^{\perp}(\mathbf{r})$

In the case when, with anisotropic distribution of defect-induced random local fields directions in the $n$-dimensional space of the order parameter, the following equality asserts

$$\langle h_{l1}^2 \rangle = \langle h_{l2}^2 \rangle = \cdots = \langle h_{ln}^2 \rangle, \tag{30}$$

the effective anisotropy quadratic in $\mathbf{h}^{\perp}(\mathbf{r})$ does not arise in the system. An example of such a distribution of defect-induced random local fields is the distribution at which the defect-induced fields are with equal probability directed collinearly to $n$ mutually perpendicular directions in the order parameter space, the latter will be chosen as the axes of Cartesian coordinate system.

In that instance the effective anisotropy can be obtained by substituting Eq. (13) for $\mathbf{s}^{\perp}(\mathbf{r})$ into the second summand in the right-hand side of Eq. (5). The corresponding inhomogeneous-exchange energy $W_{ex}^{(4)}$ takes the form

$$W_{ex}^{(4)} = -\frac{Dd}{2N^4} \int d^d \mathbf{r} \sum_{\mathbf{k}_1,\mathbf{k}_2,\mathbf{k}_3,\mathbf{k}_4} k_{1x} k_{2x} \chi^{\perp}(\mathbf{k}_1) \chi^{\perp}(\mathbf{k}_2) \chi^{\perp}(\mathbf{k}_3) \chi^{\perp}(\mathbf{k}_4) \times$$
$$\sum_{l,m,q,p} ([\mathbf{h}_l - \mathbf{s}_0(\mathbf{s}_0\mathbf{h}_l)][\mathbf{h}_q - \mathbf{s}_0(\mathbf{s}_0\mathbf{h}_q)])([\mathbf{h}_m - \mathbf{s}_0(\mathbf{s}_0\mathbf{h}_m)][\mathbf{h}_p - \mathbf{s}_0(\mathbf{s}_0\mathbf{h}_p)]) \exp\left[i\left(\mathbf{k}_1(\mathbf{r}-\mathbf{r}_l) + i\mathbf{k}_2(\mathbf{r}-\mathbf{r}_q) + \mathbf{k}_3(\mathbf{r}-\mathbf{r}_m) + \mathbf{k}_4(\mathbf{r}-\mathbf{r}_p)\right)\right]. \tag{31}$$

The integration yields $V\delta_{\mathbf{k}_4,-\mathbf{k}_1-\mathbf{k}_2-\mathbf{k}_3}$. In the end one has

$$W_{ex}^{(4)} = -\frac{Db^d d}{2N^3} \sum_{\mathbf{k}_1,\mathbf{k}_2,\mathbf{k}_3} k_{1x} k_{2x} \chi^{\perp}(\mathbf{k}_1) \chi^{\perp}(\mathbf{k}_2) \chi^{\perp}(\mathbf{k}_3) \chi^{\perp}(-\mathbf{k}_1-\mathbf{k}_2-\mathbf{k}_3) \times$$
$$\sum_{l,m,q,p} ([\mathbf{h}_l - \mathbf{s}_0(\mathbf{s}_0\mathbf{h}_l)][\mathbf{h}_q - \mathbf{s}_0(\mathbf{s}_0\mathbf{h}_q)])([\mathbf{h}_m - \mathbf{s}_0(\mathbf{s}_0\mathbf{h}_m)][\mathbf{h}_p - \mathbf{s}_0(\mathbf{s}_0\mathbf{h}_p)]) \exp[-i(\mathbf{k}_1\mathbf{r}_l + \mathbf{k}_2\mathbf{r}_q + \mathbf{k}_3\mathbf{r}_m - (\mathbf{k}_1+\mathbf{k}_2+\mathbf{k}_3)\mathbf{r}_p)]. \tag{32}$$

Let us now turn to averaging over defect-induced random fields. The result of averaging differs from zero if $l = m = q = p$ or if the indices coincide in pairs (e.g., $l = q$, $m = p$). In the latter case, the result of system energy averaging subject to the condition given by Eq. (30) appears to be independent of the mean order parameter direction. Therefore we restrict our consideration to the summand with four coinciding indices. For this summand, the exponential function in the right-hand side of Eq. (32) equals unity, and the sum over defects takes the form

$$\sum_l ([\mathbf{h}_l - \mathbf{s}_0(\mathbf{s}_0\mathbf{h}_l)][\mathbf{h}_l - \mathbf{s}_0(\mathbf{s}_0\mathbf{h}_l)])([\mathbf{h}_l - \mathbf{s}_0(\mathbf{s}_0\mathbf{h}_l)][\mathbf{h}_l - \mathbf{s}_0(\mathbf{s}_0\mathbf{h}_l)]) =$$
$$\sum_l [(\mathbf{h}_l)^2 - (\mathbf{s}_0\mathbf{h}_l)^2]^2 = \sum_l [(\mathbf{h}_l)^4 - 2(\mathbf{h}_l)^2(\mathbf{s}_0\mathbf{h}_l)^2 + (\mathbf{s}_0\mathbf{h}_l)^4]. \qquad (33)$$

If the defect-induced local fields are with equal probability directed collinearly to all coordinate axes in the order parameter space, then the quantities $\langle \mathbf{h}_l^2 h_{lj}^2 \rangle$ are equal for all $j=1, \dots, n$, and the summand $2(\mathbf{h}_l)^2(\mathbf{s}_0\mathbf{h}_l)^2$ in the right-hand side of Eq. (33) does not give rise to effective anisotropy in the order parameter space. The summand

$$\sum_l (\mathbf{s}_0\mathbf{h}_l)^4 = N x \sum_{j=1}^n s_{0j}^4 \langle h_{lj}^4 \rangle, \qquad (34)$$

is responsible for arising effective anisotropy specific for cubic crystals. The sum in the right-hand side of Eq. (34) is taken over vector components in the order parameter space.

In the case in question the quantities $\langle h_{lj}^4 \rangle$ are identical for all $j$ and $\langle h_{lj}^4 \rangle = \frac{1}{n} \langle \mathbf{h}_l^4 \rangle$, a $\sum_{j=1}^n s_{0j}^4 \langle h_{lj}^4 \rangle = \frac{1}{n} \langle \mathbf{h}_l^4 \rangle \sum_{j=1}^n s_{0j}^4$. Maximum value of the sum $\sum_{j=1}^n s_{0j}^4$ equals 1 if vector $\mathbf{s}_0$ is parallel to an axis of the Cartesian coordinate system in the order parameter space, and its minimum value is $1/n$ when this vector is directed along one of the main diagonals of the given system.

As the result the anisotropy energy density can be written as

$$w_{an} = \beta \sum_{j=1}^n s_{0j}^4, \qquad (35)$$

where the constant $\beta > 0$ is

$$\beta = -\frac{b^{2d-6}d\langle \mathbf{h}_l^4\rangle}{2nJ^3}\int\frac{k_{1x}k_{2x}d^d\mathbf{k}_1 d^d\mathbf{k}_2 d^d\mathbf{k}_3}{(2\pi)^{3d}(\mathbf{k}_1)^2(\mathbf{k}_2)^2(\mathbf{k}_3)^2(\mathbf{k}_1+\mathbf{k}_2+\mathbf{k}_3)^2}, \quad (36)$$

and integration with respect to each wave vector is performed over the Brillouin zone. For coordinate space dimension 2<$d$<4, the integral has no singularity at small $\mathbf{k}$. By introducing a dimensionless variable $\mathbf{y}=b\mathbf{k}/\pi$, we get

$$\beta = -\frac{d\langle \mathbf{h}_l^4\rangle}{(2)^{3d+1}\pi^6 nJ^3 b^d}\int\frac{y_{1x}y_{2x}d^d\mathbf{y}_1 d^d\mathbf{y}_2 d^d\mathbf{y}_3}{(\mathbf{y}_1)^2(\mathbf{y}_2)^2(\mathbf{y}_3)^2(\mathbf{y}_1+\mathbf{y}_2+\mathbf{y}_3)^2}, \quad (37)$$

integration with respect to each $\mathbf{y}$ is performed over $d$-dimensional cube with the ribs equal to 2 and parallel to coordinate axes, the cube centre is placed in the origin of coordinates. For $d = 3$, the integral in the right-hand side of Eq. (37) equals approximately $-200$.

Since $\beta > 0$, in the equilibrium state vector $\mathbf{s}_0$ is directed along one of the main diagonals of the Cartesian coordinate system in the order parameter space.

For $d = 3$, the evaluation of the effective anisotropy constant $K_{eff}$ by the order of magnitude brings the value

$$K_{eff} \sim 10^{-3}\frac{x\langle \mathbf{h}_l^4\rangle}{J^3}. \quad (38)$$

Compared to the case of collinear orientation of random field vectors, the effective anisotropy constant contains a small parameter $\langle \mathbf{h}_l^2\rangle/J^2$, it seeming natural because the energy expansion in terms of even powers of random field is performed precisely in this dimensionless parameter. A number of the order parameter components ($n = 2$ for the *X-Y* model or $n = 3$ for the Heisenberg model) is not of first importance because, in distinction to the anisotropy quadratic in $\mathbf{h}_l$, the anisotropy of fourth and higher orders in $\mathbf{h}_l$ induces the occurrence of the easy axes but not easy planes.

For the coordinate space dimensionality $d$=2, the integral in the right-hand side of Eq. (36), has a logarithmic singularity at small $\mathbf{k}$. Hence, as in the preceding section, the problem should be solved in a self-consistent way starting from the assumption of the availability of the order parameter induced by the effective anisotropy whose value can be obtained from the self-consistency equation (19). Using Eq. (20) we find

$$K_{eff} \sim \frac{x \langle \mathbf{h}_l^4 \rangle}{J^3} \ln \frac{J}{K_{eff}}. \tag{39}$$

By the first iteration we have

$$K_{eff} \sim \frac{x \langle \mathbf{h}_l^4 \rangle}{J^3} \ln \frac{J^4}{x \langle \mathbf{h}_l^4 \rangle}. \tag{40}$$

The summands containing higher powers in $\mathbf{h}_l$ can be found similarly.

## V. Phase diagram of the system

As evidenced by the foregoing, anisotropic distribution of defect-induced random local field directions gives rise to anisotropy of both "easy axis" and "easy plane" types. The Imry-Ma inhomogeneous state is suppressed only by the "easy axis" type anisotropy [2]. Therefore let us consider in some detail the situation arising at the "easy-plane" effective anisotropy type.

Gaining a tackling the question if there arises in the system the long-range order with vector $\mathbf{s}_0$ in the easy plane or the Imry-Ma inhomogeneous state, one should project all random field vectors onto the given $m$-dimensional ($n>m\geq 2$) hyperplane in the order-parameter space and treat the problem at this hyperplane. When the "easy plane" anisotropy arises, the operation should be repeated. As a result we arrive at three possible cases:

(1) Projections of random fields on the easy plane equal zero. The system behavior therewith is analogous to that of the pure system with the number of the order parameter components corresponding to the hyperplane dimensionality. In any event the Imry-Ma inhomogeneous state does not occur, though random fields induce the order parameter components normal to the easy plane.

(2) The "easy axis" anisotropy takes place in the easy plane itself. Then the problem reduces to that with the given anisotropy, but the number of the order parameter components equals *m*. Such a subject will be dealt with further.

(3) The distribution of defect-induced random field projections on the easy plane is perfectly isotropic. In this case the Imry-Ma theorem is true.

In order to understand if the Imry-Ma inhomogeneous state is realized in the system with the "easy axis" effective anisotropy type, it is necessary that the effective anisotropy constant be confronted with its critical value, wherein the state in question is suppressed. Indeed, to follow the random field fluctuations, the order parameter has to deviate from the direction corresponding to the anisotropy energy minimum. This leads to an increase in the anisotropy energy. When such a growth is not compensated by the gain in energy due to the order parameter alignment with the random field fluctuations, the Imry-Ma inhomogeneous state becomes energetically unfavorable, and the system goes back to the state with the long-range order. The requisite critical value was found in Ref. [2]

$$K_{cr} \sim J \left[\frac{x \langle \mathbf{h}_l^2 \rangle}{J^2}\right]^{\frac{2}{4-d}}. \tag{41}$$

As seen from Sections 3 and 4, the effective anisotropy induced by random defect fields is proportional to defect concentration *x*. At the same time, for space dimensionality $2 < d < 4$, the quantity $K_{cr}$ contains a higher power of defect concentration. In particular, for $d = 3$, one has $K_{cr} \propto x^2$. This suggests that in the limit of low concentration $x \to 0$ the effective anisotropy arising in any order to $\mathbf{h}_l$, is bound to exceed its critical value.

If $d = 2$, one has $K_{eff} \propto -x \ln x$, that is, in the region of small concentration the value of $K_{eff}$ also exceeds its critical value $K_{cr} \propto x$.

Thus the Imry-Ma theorem does not hold at arbitrarily small effective anisotropy of the "easy axis" type induced by defects of "random local field" type. By comparing Eqs. (24), (25) and (41), it is possible to be certain that in the case of strongly anisotropic distributions of random fields inducing the "easy axis" type anisotropy, the Imry-Ma state does not occur in the whole range of defect concentrations $x < 1$.

For slightly anisotropic distributions of random fields, the condition $K_{eff} < K_{cr}$ imposes a lower limit on the concentration of defects at which the Imry-Ma inhomogeneous state takes place. For instance, at $d = 3$ and the random local field distribution given by Eq. (26), we get the inequality

$$x > 0.1\varepsilon \frac{J^2}{\langle \mathbf{h}_l^2 \rangle}. \tag{42}$$

At $J^2/\langle \mathbf{h}_l^2 \rangle \sim 100$ and $\varepsilon \sim 10^{-3}$ we have $x > 0.01$.

The characteristic phase diagram of the system is displayed in Fig. 1.

## VI. Conclusions

An anisotropic distribution of defect-induced random local field directions initiates the effective anisotropy of either "easy axis" or "easy plane" type in the order parameter space.

The Imry-Ma theorem stating that at space dimensions $d < 4$ the introduction of an ***arbitrarily small concentration*** (italicized by the present authors) of defects of the "random local field" type in a system with continuous symmetry of the *n*-component vector order parameter (*O(n)* model) leads to the long-range order collapse and to the occurrence of a disordered state, breaks down due to existence of the "easy axis" anisotropy induced by the defects designed initially for breaking down the long-range order.

In the case of slightly anisotropic distribution of the fields, there exists a critical concentration of defects, if exceeded, the Imry-Ma inhomogeneous state can exist as an equilibrium one.

In the case of strongly anisotropic distribution of the fields, the Imry-Ma inhomogeneous state is completely suppressed and the state with the long-range ordering is realized at any defect concentration.

# Figure caption

Phase diagram of the system in variables "inverse constant of the "easy axis" type effective anisotropy $K_{eff}^{-1}$ - concentration of defects $x$ ": LRO is the phase with the long-range order, I-M is the Imry-Ma inhomogeneous phase.

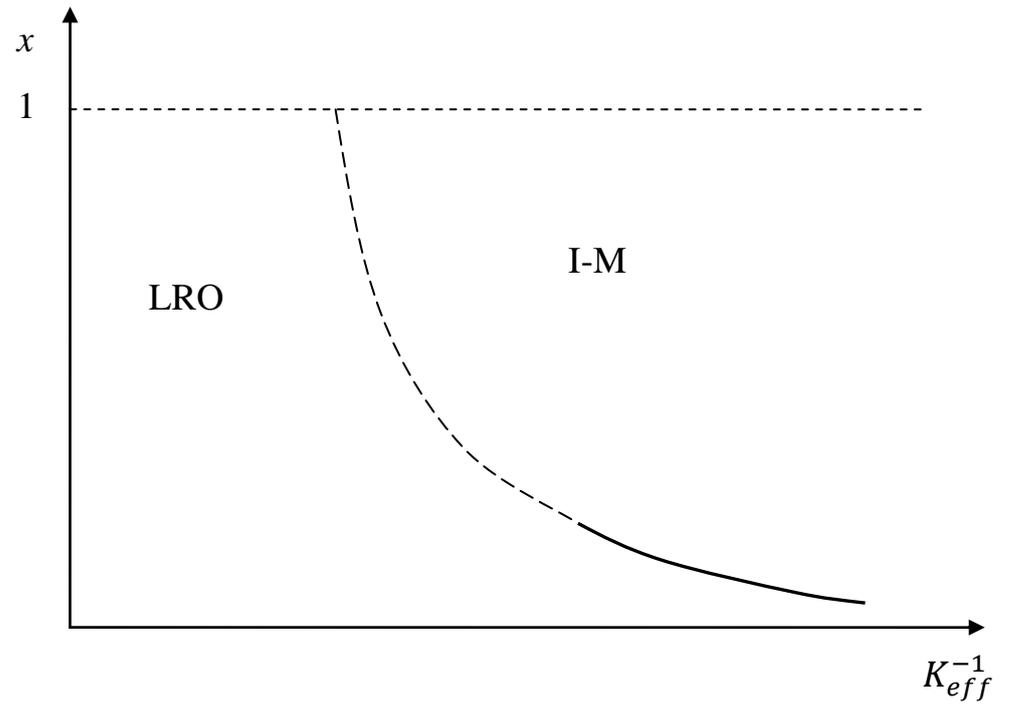